\documentclass{article}

\usepackage{psfrag}
\usepackage{amssymb}
\usepackage{amsopn}
\usepackage{amsthm}
\usepackage{lipsum}
\usepackage{amsfonts}
\usepackage{graphicx}

\usepackage{amsmath}
\usepackage{subfigure}

\newtheorem{Theorem}{Theorem}
\newtheorem{Definition}{Definition}
\newtheorem{Lemma}{Lemma}

\newcommand{\sM}{\mathcal{M}}

\newcommand{\sN}{\mathcal{N}}

\newcommand{\abs}[1]{\left\vert #1 \right\vert}
\newcommand{\norm}[1]{\left\Vert #1 \right\Vert}

\newcommand{\R}{{\mathbb R}}  %ams bold
\newcommand{\TheTitle}{On noise-induced synchronization and consensus of nonlinear network systems under input disturbances}

\title{{\TheTitle}}

\author{
 Giovanni Russo \thanks{IBM Research Ireland, Optimization, Control and Decision Science Group, grusso@ie.ibm.com.}
  \and
  Robert Shorten \thanks{IBM Research Ireland, Optimization, Control and Decision Science Group and University College Dublin,  School of Electrical and Electronic Engineering.}
  }

\begin{document}

\maketitle

\begin{abstract}
This paper is concerned with the study of synchronization and consensus phenomena in complex networks of diffusively-coupled nodes subject to external disturbances. Specifically, we make use of stochastic Lyapunov functions to provide conditions for synchronization and consensus for networks of nonlinear, diffusively coupled nodes, where noise diffusion is not just additive but it depends on the nodes' state. The sufficient condition we provide, wich links together network topology, coupling strength and noise diffusion, offers two interesting interpretations. First, as suggested by {\em intuition}, in order for a network to achieve synchronization/consensus, its nodes need to be sufficiently well connected together. The second implication might seem, instead, counter-intuitive: if noise diffusion is {\em properly} designed, then it can drive an unsynchronized network towards synchronization/consensus. Motivated by our current research in Smart Cities and Internet of Things, we illustrate the effectiveness of our approach by showing how our results can be used to control certain collective decision processes.


{\bf Notes:}
Preprint submitted to SIAM SICON.
\end{abstract}

\section{Introduction}
Over the past years, the study of synchronization and consensus of multiple interconnected systems has received considerable attention from both the Control and Physics communities, \cite{Cor_Kat_Mot_13}, \cite{Liu_Bar_Slo_11}, \cite{Che_13}. This interest has been primarily motivated by the many potential applications of synchronization and consensus, which span from rendezvous \cite{Rus_diB_d} to distributed optimization \cite{Dro_Kaw_Ege_14}, \cite{Stu_Sho_12}, power networks \cite{Dor_Bul_14} and biochemical systems analysis/control \cite{Rus_diB_09b}.

Over the past few years, several sufficient conditions have been devised ensuring, under different technical assumptions and using different approaches, synchronization/consensus of complex networks. Examples of such approaches include passivity-based techniques \cite{Bur_DeP_15}, input-output stability \cite{Sca_Arc_Son_10}, Lyapunov techniques \cite{jita06}, convergent systems \cite{Pav_Pog_Wou_Nij}, incremental stability, \cite{For_Sep_12}, contraction theory \cite{Rus_diB_Son_13} and the Master Stability Function(mainly used within the Physics Community) \cite{Pec_Car_Joh_Mar_97}. An assumption that is often made in Literature is that the network of interest is {\em noise-free}. This assumption is not realistic for most real world applications of synchronization and consensus, where noise plays a key role in destroying or generating those coordinated behaviors. A class of networks where noise cannot be neglected is the one where the network nodes {\em interact} with the environment. Such networks are relevant to many applications in Nature and Technology, with examples ranging from synchronization of biochemical reactions to coordination of smart devices connected over the Internet of Things (IoT).

This paper is concerned with the study of synchronization and consensus in networks where nodes interact with the external environment. This interaction results in a perturbation of the nodes' intrinsic dynamics by some external (white) {\em noise}. Specifically, we will consider networks where nodes' dynamics are nonlinear and where the diffusion of noise depends on the nodes' state. By using stochastic Lyapunov techniques \cite{Mao_97}, we present a sufficient condition ensuring that the network synchronizes (or achieves consensus) in the presence of the external noise. We will also show that our condition, which links together network topology, coupling strength, node and diffusion dynamics, has to two interesting interpretations. First, as expected, the condition implies that in order for the network to achieve synchronization/consensus, the nodes need to be {\em strongly connected} together so as to overcome a {\em threshold} generated by noise. On the other hand, our result also offers a perhaps counter-intuitive interpretation as it shows that, under certain assumptions, noise drives the network to achieve synchronization and consensus. In turn, this implies that, if properly {\em designed}, noise can be turned into a {\em distributed control input} to induce synchronization/consensus. 

The paper is organized as follows. We start with introducing the notation and mathematical preliminaries in Section \ref{sec:math_prel}. Then, in Section \ref{sec:probl_stat}, we formalize the problem statement while, in Section \ref{sec:net_synch}, we provide a sufficient condition for network synchronization/consensus. Finally, in Section \ref{sec:applications}, we apply our result to study collective decision processes. 

\subsection*{Related Work}

The problem of analyzing/controlling synchronization and consensus in continuous-time networks affected by noise is attracting many researchers. For example, recently, in \cite{Tan_Li_15}, the consensus problem has been studied for a network of integrators when the information exchanged between nodes is corrupted by some white noise. Also, in  \cite{Sri_Leo_14}, a similar model (i.e. the so-called drift-diffusion model, {\em DDM}) is used to study the dynamics of collective decision processes. In order to devise their results, in such papers it is assumed that the noise affecting each node is additive with constant diffusion rate and that the dynamics at the network nodes are linear $1$-dimensional systems or integrators. Recently, the assumptions on the nodes' dynamics have been relaxed in \cite{Wel_Kat_Mot_15}. Specifically, in this paper, an approach to study synchronization in networks of nonlinear nodes subject to additive noise has been presented. However, in order to obtain their results, the authors recast the synchronization problem as a stochastic optimization problem and then apply numerical methods to solve it. Interestingly, a large body of literature is emerging which is devoted to study network dynamics when the external disturbance acting on the nodes is not a white noise but it is rather the output of some exogenous system, see e.g. \cite{DeP_Jay_14} and \cite{Wei_Van_13}. In this case, as shown in e.g. \cite{Kim_DeP_15} (see also references therein), nonlinear nodes' dynamics can be considered, with the external disturbance affecting the dynamics via some possibly nonlinear function. 

\section{Mathematical Preliminaries}\label{sec:math_prel}

\subsection{Notation}
In this paper, we will denote with $I_n$ the $n\times n$ identity matrix and with $1_{n\times m}$ the $n \times m$ matrix having all of its elements equal to $1$. The vector/matrix Frobenius norm will be denoted with $\norm{\cdot}_F$ and the vector/matrix Euclidean norm will be denoted with $\abs{\cdot}$. The trace of a square matrix, say $A$, will be denoted with $tr\left\{A\right\}$. Finally, we will denote by $\sM$ the $n$-dimensional subspace spanned by the vector $1_n$.

\subsection{Stochastic differential equations}
Consider an $n$-dimensional stochastic differential equation of the form
\begin{equation}\label{eqn:ito_gen}
dx = f(t,x)dt + g(t,x)dB,
\end{equation}
where: (i) $x \in \R^n$ is the state variable; (ii) $f:\R^+\times\R^n\rightarrow\R^n$ belongs to $\mathcal{C}^2$; (iii) $g:\R^+\times\R^n\rightarrow\R^{n\times d}$ belongs to $\mathcal{C}$; (iv) $B= [B_1,\ldots,B_d]^T$ is a $d$-dimensional Brownian motion. Throughout this paper we will assume that both $f$ and $g$ obey the local Lipschitz condition and the linear growth condition, see e.g. \cite{Oks_07}. This implies that for any given initial condition $x(t_0) = x_0$, $t\ge 0$, equation (\ref{eqn:ito_gen}) has a unique global solution. We will also assume that $f(t,0) = g(t,0) = 0$ and the solution $x=0$ will be said the {\em trivial solution} of (\ref{eqn:ito_gen}).

Following \cite{Kar_93}, \cite{Roh_76}, we say that a sequence of stochastic variables, $\left\{V_1,V_2,\ldots\right\}$ converges almost surely (a.s.) to the stochastic variable $V$ if
$$
\mathbb{P}\left(\left\{w: \lim_{n\rightarrow +\infty}V_n(w)=V(w)\right\}\right) = 1.
$$
That is, the sequence converges to $V$ with probability $1$. We are now ready to give the following definition which characterizes stability of the trivial solution, see \cite{Mao_97}. 
\begin{Definition}
The trivial solution of (\ref{eqn:ito_gen}) is said to be almost surely exponentially stable if for all $x\in\R^n$, $\lim_{t\rightarrow +\infty}\sup\frac{1}{t}\log\left(\abs{x(t)}\right) <0$, $a.s.$.
\end{Definition}

Let $V(t,x):\R^+\times\R^n\rightarrow\R^+$, $V(t,x) \in \mathcal{C}^{1\times2}$, i.e. $V(t,x)$ is twice differentiable with respect to $x$ and differentiable with respect to $t$. By the Ito formula we have:
$$
dV(t,x) = LV(t,x)dt +V_x(t,x)g(t,x)dB,
$$
where: (i) $LV(t,x) = V_t(t,x) + V_x(t,x)f(t,x) + \frac{1}{2}tr\left\{ g(t,x)^TV_{xx}g(t,x)(t,x)\right\}$;(ii) $V_x = \left[V_{x_1},\ldots,V_{x_n}\right]$; (iii) $V_{xx}$ is the $n\times n$ dimensional matrix having as element $ij$ $V_{x_ix_j}$ (where $V_{x_i} := \partial V(t,x)/\partial x_i$ and $V_{x_ix_j} := \partial^2 V(t,x)/\partial x_j \partial x_i$). In the rest of the paper we assume that the diffusion function $g(t,x)$ is linear in $x$. The following result from \cite{Mao_97} provides a sufficient condition for the trivial solution of (\ref{eqn:ito_gen}) to be almost surely exponentially stable.
\begin{Theorem}\label{thm:stability_ito}
Assume that there exists a non-negative function $V(t,x) \in \mathcal{C}^{1\times2}$ and constants $p>0$, $c_1>0$, $c_2\in\R$, $c_3\ge 0$, such that $\forall x \ne 0$ and $\forall t \in \R^+$: ({\bf H1}) $c_1\abs{x}^p \le V(t,x)^p$; ({\bf H2}) $LV(t,x) \le c_2 V(t,x)$; ({\bf H3}) $\abs{V_x(t,x)g(t,x)}^2 \ge c_3 V(t,x)^2$.
Then:
$ \lim_{t\rightarrow +\infty}\sup\frac{1}{t}\log\left(\abs{x(t)}\right) \le -\frac{c_3-2c_2}{p}$, $a.s.$. In particular, if $c_3>2c_2$, then the trivial solution of (\ref{eqn:ito_gen}) is almost surely exponentially stable. 
\end{Theorem}

\subsection{Complex networks}
Throughout this paper, we will consider systems interacting over some graph, $\mathcal{G} = (\mathcal{V},\mathcal{E})$, where $\mathcal{V}$ is the set of vertices (or nodes in what follows) and $\mathcal{E}$ is the set of edges. We assume all the graphs in this paper are undirected and denote the edge between node $i$ and node $j$ as $(i,j)$. We will denote with $\sN_i$ the set of neighbors of node $i$, i.e. $\sN_i :=\left\{j:(i,j)\in\mathcal{E}\right\}$. Let $N$ be the number of nodes in the network. Then (see e.g. \cite{God_Roy_01}) the Laplacian matrix associated to the graph, $L$, is the $N\times N$ symmetric matrix defined as $L = \Delta - A$, where: (i) $A$ is the adjacency matrix of $\mathcal{G}$; (ii) $\Delta$ is the graph degree matrix. The following Lemma from \cite{Hor_Joh_99} will be used in this paper.

\begin{Lemma}\label{lem:laplacian}
Denote with $L$ the Laplacian matrix of an undirected network. The following properties hold: (i) $L$ has a simple zero eigenvalue and all the other eigenvalues are positive if and only if the network is connected; (ii) the eigenvector associated to the zero eigenvalue is $1_N$, i.e. the $N$-dimensional vector having all of its elements equal to $1$; (iii) the smallest nonzero eigenvalue, $\lambda_2$, satisfies $\lambda_2 = \min_{v^T 1_N=0, v \ne 0}\frac{v^TLv}{v^Tv}$.
\end{Lemma}
In the rest of this paper, $\lambda_2$ will be termed as the graph algebraic connectivity. 

\section{Problem statement}\label{sec:probl_stat}
Throughout this paper, we will consider stochastic networks described by the following stochastic differential equation:
$$
dx_i = \left[f(t,x_i) + \sigma\sum_{j\in\sN_i}\left(x_j-x_i\right)\right]dt + g(t,x_i)db, \ \ \ i = 1,\ldots,N,
$$
where: (i) $x_i \in \R^n$; (ii) $f(t,x_i):\R^+\times\R^n\rightarrow\R^n$ is the smooth {\em nominal} nodes dynamics; (iii) $\sigma$ is the coupling strength; (iv) $b(t)\in\R$ is the standard Brownian process describing the environmental noise acting on the network; (v) $g(t,X): \R^+\times\R^{nN}\rightarrow\R^{n}$ is the smooth $n$-dimensional vector modeling how the environmental noise {\em diffusesn} to the network. Note that network dynamics can be written in compact form as follows:
\begin{equation}\label{eqn:network}
dx = \left[F(t,X) - \sigma (L\otimes I_n)X\right]dt + G(t,X)db,
\end{equation}
with:
\begin{itemize}
\item $x = [x_1^T,\ldots,x_N^T]^T$;
\item $F(t,X) = [f(t,x_1)^T,\ldots,f(t,x_N)^T]^T$;
\item $G(t,X) = [g(t,x_1)^T,\ldots,g(t,x_N)^T]^T$.
\end{itemize}

The goal of this paper is to address the so-called {\em synchronization problem}. This is formalized with the following definition.
\begin{Definition}
Let $s(t) = \frac{1}{N}\sum_{j=1}^N x_j(t)$. We will say that network (\ref{eqn:network}) achieves stochastic synchronization if
$$
\lim_{t\rightarrow +\infty}\sup\frac{1}{t}\log\left(\abs{x_i(t)-s(t)}\right) <0 \ \ \ a.s. \ \ \ \forall i=1,\ldots,N.
$$
\end{Definition}
That is, stochastic synchronization essentially means that all the network nodes converge towards a synchronous solution that {\em averages} the noise diffusion within the network.  Also, note that in the case where nodes' dynamics are integrator dynamics, then the definition of stochastic synchronization simply becomes a definition for consensus.

\subsection*{A discussion on the model}
\begin{figure}[thbp]
\begin{center}
  \includegraphics[width=10cm]{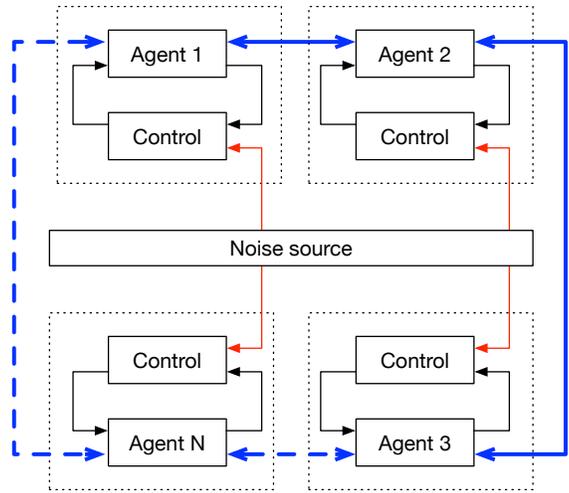}
  \includegraphics[width=10cm]{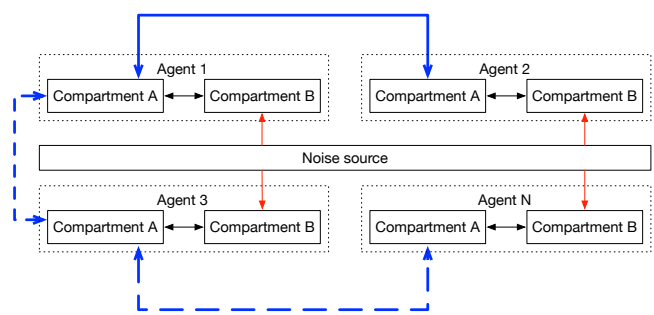}
  \caption{The set-ups motivating the class of equations studied in this paper. {\bf Top panel:} a number of locally controlled interconnected devices, with the local controller collecting data from the external environment. {\bf Bottom panel:} a network of compartmental systems, where a subset of the compartments (i.e. {\em compartment B}) is affected by an external (typically, environmental) noise.}
  \label{fig:model}
  \end{center}
\end{figure}

Equations similar to (\ref{eqn:network}) naturally arise when modeling networked systems from IoT and biochemical applications. For example, as outlined in Figure \ref{fig:model} (top panel), IoT applications are typically deployed by interconnecting a number of {\em smart} devices. Each device of the networked system consists of some intrinsic {\em agent} dynamics, which is controlled by a local control algorithm. Typically, the control algorithm also takes information/data from the external environment and this leads the controller to be affected by an environmental noise. This is the case of many complex cyber-physical systems, \cite{IEEE_2012}, with applications ranging from the coordination of team of quadricopters, \cite{Heh_Dan_14}, to supervisory human control, \cite{Pet_Sri_Tay_Sur_Eck_Bul_15} and decision-making processes, \cite{Sri_Leo_14}. Another important instance where equations similar to (\ref{eqn:network}) arise is when the agents of a networked system are made up by a number of compartments (see the bottom panel of Figure \ref{fig:model}), with one of the compartments being affected by an external noise. These {\em compartmental} models are widely used in pharmacology and many other biochemical applications. For examples, for a biochemical network, each of the compartments might be a subset of biochemical reactions or even macro-systems, like an organ or blood, \cite{Sha_Arc_15}. 

The two classes of networked systems described above can be recast into equations of the form (\ref{eqn:network}). Specifically, the function $f(t,x_i)$ would model the noise free component of the $i$-th agent in a networked system (i.e. the dynamics of a device or the noise-free compartment ofa compartmental system), while the function $g(x_i)$ would model the sub-components of the nodes' dynamics which are affected by the external noise (i.e. the $i$-th local controller taking inputs from the external world or a compartment of a compartmental system which is affected by environmental noise).

\section{A sufficient condition for stochastic synchronization}\label{sec:net_synch}

Let $S = 1_n\otimes s(t)$. The following result provides a sufficient condition ensuring stochastic synchronization of network (\ref{eqn:network}).
\begin{Theorem}\label{thm:network_nodes}
Assume that the following conditions hold for network (\ref{eqn:network}):
\begin{enumerate}
\item there exists some constant, say $K_f$, such that
$$
(X-Y)^T\left[F(t,X)-F(t,Y)\right]\le K_f(X-Y)^T(X-Y),
$$,
$\forall X,Y \in \R^{Nn}$, $\forall t\in\R^+$;
\item there exists some constant, say $K_G$, such that 
$$
\abs{G(t,X)- G(t,Y)} \le K_G\abs{X-Y},
$$ 
$\forall X,Y \in \R^{Nn}$, $\forall t\in\R^+$;
\item there exists some constant, say $\bar K_G$, such that 
$$
\abs{(X-Y)^T(G(t,X)- G(t,Y))}^2 \ge \bar K_G\abs{X-Y}^4,
$$
$\forall X,Y \in \R^{Nn}$, $\forall t\in\R^+$;
\item $\sigma\lambda_2 > K_f + \frac{K_g^2- 2\bar K_g^2}{2}$.
\end{enumerate}
Then, (\ref{eqn:network}) achieves stochastic synchronization.
\end{Theorem}
\proof
Let $s(t) = \frac{1}{N}\sum_{j=1}^Nx_j(t)$. We have $ds = \frac{1}{N}\sum_{j=1}^Ndx_j,$ and therefore:
$$
ds = \frac{1}{N}\sum_{j=1}^Nf(t,x_j)dt+\frac{1}{N}\sum_{j=1}^Ng(t,x_j)dB.
$$
Since by assumptions the function $g(t,x)$ is linear in $x$, we have that:
$$
\frac{1}{N}\sum_{j=1}^Ng(t,x_j) = g(t,\sum_{j=1}^N\frac{1}{N}x_j) = g(t,s).
$$
Thus, in compact form:
\begin{equation}\label{eqn:sync_sol}
dS = \frac{1}{N}(1_{N\times N}\otimes I_n)F(t,X)dt + G(t,S).
\end{equation}
From (\ref{eqn:network}) and (\ref{eqn:sync_sol}) we can then get the stochastic differential equation for $e = X - S$. Namely:
\begin{equation}\label{eqn:error}
de = \left[\tilde F(t,e)\right]dt + \left[\tilde G(t,e)\right]db,
\end{equation}
where:
\begin{itemize}
\item $\tilde F(t,e) =F(t,e+S) - \sigma LX - \frac{1}{N}(1_{N\times N}\otimes I_n)F(t,e+S)$;
\item $\tilde G(t,e) = G(t,e+S) - G(t,S)$. 
\end{itemize}
Now, note that $e=0$ is the trivial solution for (\ref{eqn:error}). In fact:
$$
F(t,S) - \frac{1}{N}1_{N\times N}F(t,S) = 0,
$$
and, at the same time, $\tilde G(t,0) = 0$. This means that we can use Theorem \ref{thm:stability_ito} to show network synchronization. To this aim, let $V(t,e) = V(e) = \frac{1}{2}e^Te$, from Theorem \ref{thm:stability_ito} we need to show that there exists $c_2\in\R$, $c_3\ge 0$, such that $c_3 > 2c_2$. In order to prove this, we will now estimate $LV(e)$ and $\abs{V_e(t,e)\tilde G(t,e)}^2$.

In order to compute the term $LV(e)$, first note that $V_t(e) = 0$. Let's now compute the term $V_e(e)\tilde F(t,e)$. We have:
$$
V_e(e)\tilde F(t,e) = e^T\left[F(t,e+S) - \sigma (L\otimes I_n)(e+S) - \frac{1}{N}(1_{N\times N}\otimes I_n)F(t,e+S) \right],
$$
and by adding and subtracting $F(t,S) = 1_N\otimes f(t,s)$ we get
$$
\begin{aligned}
V_e(e)\tilde F(t,e) = \\
= e^T\left[F(t,e+S) + F(t,S) - F(t,S) - \sigma (L\otimes I_n)(e+S) + \right.\\
\left. - \frac{1}{N}(1_{N\times N}\otimes I_n)F(t,e+S)\right].
\end{aligned}
$$
On the other hand, note that:
$$
e^T\left[F(t,S) - \frac{1}{N}(1_{N\times N}\otimes I_n)F(t,e+S) \right] = 0,
$$
while:
$$
- \sigma (L\otimes I_n)(e+S) = -\sigma (L\otimes I_n) e.
$$
Thus:
$$
V_e(e)\tilde F(t,e) = e^T\left[F(t,e+S)-F(t,S) - \sigma (L\otimes I_n)e \right].
$$
That is,
\begin{equation}\label{eqn:diffusion}
V_e(e)\tilde F(t,e) = e^T\left[F(t,e+S)-F(t,S)\right] - \sigma e^T(L\otimes I_n)e.
\end{equation}
Now:
$$
\begin{aligned}
V_e(e)\tilde F(t,e) \le \\
\le e^T\left[F(t,e+S)-F(t,S)\right] - \sigma min_{e\ne0}\left\{e^T(L\otimes I_n)e\right\} =\\
= e^T\left[F(t,e+S)-F(t,S)\right] - \sigma \lambda_2 e^Te,
\end{aligned}
$$
where the last equality follows from Lemma \ref{lem:laplacian}. Finally, by hypothesis $1$, we have that
\begin{equation}\label{eqn:LV}
V_e(e)\tilde F(t,e) \le \left(K_f-\sigma\lambda_2\right)e^Te = 2\left(K_f-\sigma\lambda_2\right)V(e).
\end{equation}
The next step to estimate $LV(e)$ is that of computing the term 
$$
\frac{1}{2}tr\left\{ \tilde G(t,e+S)^TV_{ee}\tilde G(t,e+S)\right\}.$$
Since $V_{ee}(e) = 1$, such a condition simply becomes:
$\frac{1}{2}tr\left\{ \tilde G(t,e+S)^T\tilde G(t,e+S)\right\}$.
Also, recall that for any matrix, say $A$, we have $\norm{A}_F^2 = tr\left\{ A^TA\right\}$ and thus:
$$
\left(tr\left\{ \tilde G(t,e+S)^TV_{ee}\tilde G(t,e+S)\right\}\right)^{1/2} =\norm{\tilde G(t,e+S)}_F = \abs{\tilde G(t,e+S)},
$$
where the last inequality follows from the fact that for any vector, say $a$, $\norm{a}_F = \abs{a}$.
Now, by hypothesis $2$, we have:
$$
\abs{\tilde G(t,e+S)} =  \abs{G(t,e+S) - G(t,S)}\le K_G\abs{e},
$$
and therefore we have:
\begin{equation}\label{eqn:VXX_2}
\begin{aligned}
\frac{1}{2}tr\left\{ \tilde G(t,e+S)^TV_{ee}\tilde G(t,e+S)\right\} = \frac{1}{2} \abs{G(t,e+S) - G(t,S)}^2 \le \\
\le \frac{1}{2} K_G^2\abs{e}^2 = K_G^2V(e).
\end{aligned}
\end{equation}

Combining (\ref{eqn:VXX_2}) and (\ref{eqn:LV}) we get:
\begin{equation}\label{eqn:c2}
LV(e) \le \left(2K_f+K_G^2-2\sigma\lambda_2\right)V(e).
\end{equation}

In order to complete the proof, we now need to compute a lower bound for $\abs{V_e(e)\tilde G(t,e+S)}$. In order to do so first note that
$$
\abs{V_e(e)\tilde G(t,e)} = \abs{e^T\left(G(t,e+S) - G(t,S)\right)}.
$$
Thus, by Hypothesis $3$, we have
\begin{equation}\label{eqn:sec_term}
\abs{V_e(e)\tilde G(t,e)}^2 \ge \bar K_g^2 \abs{e}^4 = \bar K_g \left(e^Te\right)^2 = 4\bar K_g^2V(e)^2:=c_3V(e)^2.
\end{equation}

We can then conclude the proof by noticing that, by Hypothesis $4$
$$
4\bar K_g^2 > 2\left(2K_f+K_g^2-2\sigma\lambda_2\right).
$$
Therefore, by means of Theorem \ref{thm:stability_ito}, 
$\lim_{t\rightarrow +\infty}\sup\frac{1}{t}\log\left(\abs{e(t)}\right) <0$, $a.s.$, thus proving the result. \endproof

\subsection*{Remarks on Theorem \ref{thm:network_nodes}}
\begin{itemize}
\item Hypothesis $1$ is sometimes known in the literature as {\em QUAD}. As shown in \cite{deL_diB_Rus_11}, this condition can be linked to Lipschitz and contraction conditions of the vector field, \cite{}, \cite{}. Specifically, in this latter case, this would imply $K_f < 0$;
\item Hypothesis $2$ implies that the noise diffusion is bounded. Note that, by assumption, the function $g(t,x)$ is linear in $X$ and therefore there exists some $n\times n$ matrix, say $M_g(t)$, such that $g(t,x) := M_g(t)x$. Hence, the function $G(t,X)$ can be expressed as $G(t,X) := M_G(t)X$, where $M_G(t)$ is the $nN\times nN$ block-diagonal matrix having on its main diagonal the matrices $M_g(t)$. Now, Hypothesis $2$ of Theorem \ref{thm:network_nodes} can be characterized in therms of the eigenvalues of the matrix $M_g(t)$. Specifically, let $\lambda_{\max}(M_G(t))$ and $\lambda_{\max}(M_g(t))$ be the largest (time-varying) eigenvalue of $M_G(t)$ and $M_g(t)$, respectively. Then: 
$$
\abs{G(t,X) - G(t,y)} = \abs{M_G(t)(X-Y)}\le \lambda_{\max}(M_G(t))\abs{X-Y},
$$ 
where the last inequality follows from the fact that $M_G(t)$ is a block diagonal matrix. Therefore, we have that $K_G = \max_t\left\{\lambda_{\max}(M_g(t))\right\}$; 
\item  while Hypothesis $3$ allows the noise to be {\em persistent}. Again, such a condition can be expressed in terms of the eigenvalues of $M_g(t)$. Specifically, it can be shown that in this case $\bar K_G := \min_t\left\{\lambda_{\min}(M_g(t))\right\}$, where $\lambda_{\min}(M_g(t))$ is the smaller eigenvalue of $M_g(t)$;
\item Finally, Hypothesis $4$ links together the algebraic connectivity of the network graph and coupling strength between nodes. Specifically, such a condition implies that synchronization is attained if $\sigma \lambda_2 > \tilde K := K_f + \frac{K_g^2-2\bar K_g^2}{2}$. The threshold $\tilde K$ depends on the dynamics of the node on noise. Interestingly, this hypothesis is implying that noise can potentially be helpful for synchronization. Specifically, if $2\bar K_g^2 > K_g^2$, then the effect of noise is that of lowering $\tilde K$, thus helping synchronization rather than being disruptive. We will build upon this idea in Section \ref{sec:applications}, where we will show that a common opinion among a set of agent can be forced by {\em inducing} a noise of sufficiently high intensity within the network.
\end{itemize}

\section{An application: inducing a collective decision through noise}\label{sec:applications}

The problem we consider in this Section is that of understanding whether a population of interconnected agents which are affected by some external disturbance is able to achieve a common decision. Recently, this problem has become particularly relevant in IoT applications when, based on the local collection of data (observations), a group of diffusively coupled objects, with possibly differing likelihoods, is in charge of detecting the occurrence of a given event among two alternatives. This networked system can be modeled with the following  differential equation:
$$
\dot x_i = x_i^\ast -x_i^3 + \sum_{j\in\sN_i}(x_j-x_i),
$$
and finding conditions for the onset of a collective decision  is equivalent to devising synchronization conditions for the network above. In such a network, the intrinsic dynamics of the nodes is a bistable system (i.e. $x_i^\ast-x_i^3$) and nodes are diffusively coupled. In the equation above, the intrinsic dynamics of the nodes is perturbed by some external white noise, $w(t)$ and, in particular, the term $x_i^\ast$ can be decomposed as $rx_i + \sigma_n x_iw(t)$. That is, $x_i^\ast$ consists of a noise-free component, $rx_i$, and of a term which is affected by noise and hence acts as a perturbation on the node dynamics. The equation above can be therefore recast as the following stochastic differential equation:
\begin{equation}\label{eqn:dec_making}
dx_i = \left[r x_i - x_i^3 + \sum_{j\in\sN_i}(x_j-x_i)\right]dt + \left[\sigma_n x_i\right]db.
\end{equation}

Equations similar to (\ref{eqn:dec_making}) also arise in the context of the study of human performance in a two alternative decision making process. Specifically, the formation process of an individual can be modeled by the {\em Drift Diffusion Model} (DDM), see e.g. \cite{Rat_McK_08}, \cite{Pet_Sri_Tay_Sur_Eck_Bul_15}  and references therein: $dx = \beta dt + \sigma dB$, where $\beta \in\R$ is the drift rate, $\sigma >0$ is the diffusion rate and $x(t)$ is the opinion at time $t$. Recently, in the remarkable work \cite{Sri_Leo_14} the DDM has been extended to study collective decisions. In order to do so, the authors used the coupled version of the DDM: $dx_i = \left[\beta + \sum_{j\in\sN_i}(x_j-x_i)\right]dt + \sigma_b db_i
$, where: (i) $x_i$ is the aggregate evidence of the $i$-th agent in the network; (ii) $db_i$ models the (external) noise affecting the data collected by $i$-th agent; (iii) $\sigma_b$ models the strength of noise diffusion on each network node. 

We will now use Theorem \ref{thm:network_nodes} to study the effects of noise on the capability of network (\ref{eqn:dec_making}) to achieve a common decision. The first step to apply Theorem \ref{thm:network_nodes} is to compute the parameters $K_f$, $K_G$ and $\bar K_G$.

{\bf Computation of $K_f$:} this parameter can be computed by noticing that $(x-y)(f(x) - f(y)) = (x-y)(r(x-y)-(x^3-y^3))$ and therefore $(x-y)(f(x) - f(y)) \le r(x-y)^2$. That is, $K_f = r$.

{\bf Computation of $K_G$ and $\bar K_G$:} note that $\abs{G(t,X) - G(t,y)} =\abs{M_G(X-Y)}$, where $M_G$ is the diagonal matrix having on its main diagonal the terms $\sigma_n$. Thus, we have $\abs{M_G(X-Y)} \le \sigma_n \abs{X-Y}$, i.e. $K_G = \sigma_n$. Analogously, in order to compute $\bar K_G$ it suffices to note that $\abs{(X-Y)^T(G(t,X) - G(t,Y))}= \abs{(X-Y)^TM_G(X-Y)}$ and therefore $\bar K_G = \sigma_n$.

\begin{figure}[thbp]
\begin{center}
\centering \psfrag{x}[c]{{$t$}}
\psfrag{y}[c]{{$x_i$'s}}
  \includegraphics[width=13cm]{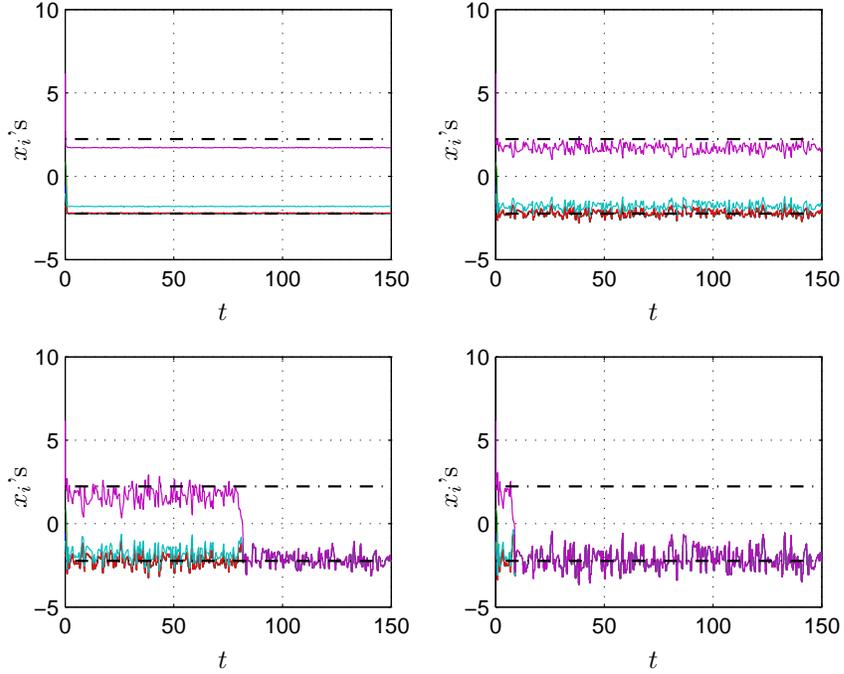}
  \caption{Time evolution for the $5$ nodes chain topology network for different values of $\sigma_n$. Specifically, in top left panel, the behavior is shown for $\sigma_n = 0.1$ while in the top right panel the time evolution is for $\sigma_n = 2$. Such values do not fulfill our condition for synchronization. In the bottom panels, instead, the time behaviors are shown for $\sigma_n = 4$ (bottom left) and $\sigma_n = 8$ (bottom right). Such values of $\sigma_n$ fulfill our condition for synchronization and in fact a collective decision is made by the network nodes. In the figure, the dashed black lines (colors online) show the stable fixed points for the nodes' intrinsic dynamics ($rx_i - x_i^3$). The initial conditions for the network nodes have been taken from the standard distribution with mean $0$ and standard deviation $5$.}
  \label{fig:decision_making_state}
  \end{center}
\end{figure}

\begin{figure}[thbp]
\begin{center}
\centering \psfrag{x}[c]{{$t$}}
\psfrag{y}[c]{{$\sigma_nx_i$'s}}
  \includegraphics[width=13cm]{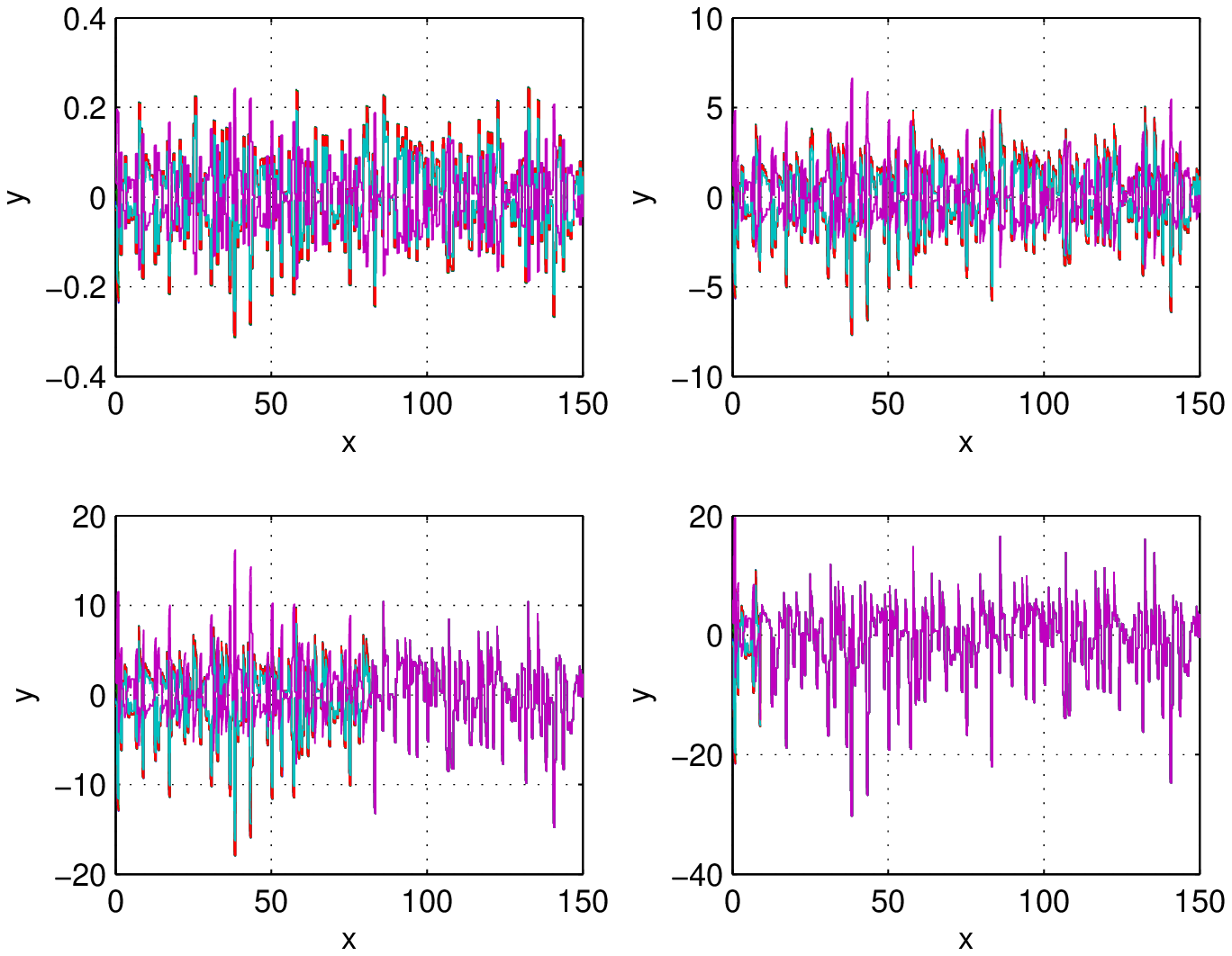}
  \caption{Time behavior for the terms $\sigma_nx_i$'s corresponding to the time evolution of the network state shown in Figure \ref{fig:decision_making_state}. In the top left panel $\sigma_n = 0.1$, while in the top right panel $\sigma_n = 2$. For the bottom panels we have $\sigma_n = 4$ (bottom left) and $\sigma_n = 8$ (bottom right).}
  \label{fig:decision_making_noise}
  \end{center}
\end{figure}

Now, following Theorem \ref{thm:network_nodes}, the network synchronizes (i.e. a common decision is achieved by the agents) if $\lambda_2 > r - \frac{\sigma_n^2}{2}$. This condition has an interesting interpretation. Specifically, it implies that a collective decision can be made by network (\ref{eqn:dec_making}) if: (i) network nodes are {\em well} connected together (i.e. $\lambda_2$ is sufficiently large); (ii) noise is sufficiently {\em strong} (i.e. $\sigma_n$ is sufficiently large). In order to provide a numerical demonstration for our synchronization condition, consider a simple network of $5$ nodes having a chain topology (i.e. $\lambda_2 = 0.38$) and where $r = 5$. Following our theoretical prediction, the network will synchronize if $\sigma_n^2/2 > r -\lambda_2$, i.e. $\sigma_n > 3$. Such a prediction is confirmed by Figure \ref{fig:decision_making_state}, where the network behavior is shown for different values of $\sigma_n$. The time behavior for the terms $\sigma_nx_i$'s is instead shown in Figure \ref{fig:decision_making_noise}.

\section{Conclusions}

In this paper we presented a new sufficient condition ensuring that a network affected by an exogenous white noise synchronizes or achieves consensus. Our results are based on the use of stochastic Lyapunov functions and this allowed us to consider networks modeled via Ito stochastic differential equations, with nonlinear nodes and with the noise diffusion depending on the nodes' state. Our result offers two interesting interpretations. First, as {\em intuition} might suggest, in order for a network to achieve synchronization/consensus, its nodes need to be sufficiently well connected together, i.e. the network needs to have a sufficiently higher algebraic connectivity and coupling strength. A second implication might seem, however, counter-intuitive. Specifically, our result implies that, if noise diffusion is {\em properly} designed, then it can drive an unsynchronized network towards synchronization/consensus.
The idea of using noise to control synchronization/consensus has been motivated by our research in smart cities and IoT, where a network of smart objects typically needs to take decisions by observing some local quantity and collecting evidences. In order to show the effectiveness of our approach, we apply our results to the study of collective decision processes. Specifically, in accordance with our theoretical predictions, we show that it is possible to force a collective decision for a network by properly inducing noise. While this is particularly relevant for IoT applications, where noise could be used to preserve privacy of a network node, this might also indicate, at a more fundamental level, that the opinion formation of a population can be driven towards a consensus by flooding the network with a sufficiently large amount of noise.


\begin{thebibliography}{10}

\bibitem{IEEE_2012}
{\em IEEE Special issue on Cyber-Physical Systems}, vol.~1000, Proceedings of
  the IEEE, 2012.

\bibitem{Bur_DeP_15}
{\sc M.~B{\"{u}}rger and C.~D. Persis}, {\em Dynamic coupling design for
  nonlinear output agreement and time-varying flow control}, Automatica, 51
  (2015), pp.~210--222.

\bibitem{Che_13}
{\sc G.~Chen}, {\em Problems and challenges in control theory under complex
  dynamical network environments}, Acta Automatica Sinica, 39 (2013),
  pp.~321--321.

\bibitem{Cor_Kat_Mot_13}
{\sc S.~Cornelius, W.~Kath, and A.~Motter}, {\em Realistic control of network
  dynamics}, Nature Communications, 4 (2913), p.~1942.

\bibitem{deL_diB_Rus_11}
{\sc P.~de~Lellis, M.~di~Bernardo, and G.~Russo}, {\em On quad, lipschitz and
  contracting vector fields for consensus and synchronization of networks},
  IEEE Transactions on Circuits and Systems I, 58 (2011), pp.~576--583.

\bibitem{Dor_Bul_14}
{\sc F.~Dorfler and F.~Bullo}, {\em Synchronization in complex networks of
  phase oscillators: a survey}, Automatica, 50 (2014), pp.~1539--1564.

\bibitem{Dro_Kaw_Ege_14}
{\sc G.~Droge, H.~Kawashima, and M.~Egerstedt}, {\em Continuous-time
  proportional-integral distributed optimization for networked systems},
  Journal of Decision and Control, 1 (2014), pp.~191--213.

\bibitem{For_Sep_12}
{\sc F.~Forni and R.~Sepulchre}, {\em A differential lyapunov framework for
  contraction analysis}.
\newblock 2012.

\bibitem{God_Roy_01}
{\sc C.~Godsil and G.~Royle}, {\em Algebraic Graph Theory}, Springer Verlag
  (New York), 2001.

\bibitem{Heh_Dan_14}
{\sc M.~Hehn and R.~D’Andrea}, {\em A frequency domain iterative learning
  algorithm for high-performance, periodic quadrocopter maneuvers},
  Mechatronics, 24 (2014), pp.~954--965.

\bibitem{Hor_Joh_99}
{\sc R.~A. Horn and C.~R. Johnson}, {\em Matrix Analysis}, Cambridge University
  Press (Cambridge, UK), 1999.

\bibitem{jita06}
{\sc G.-P. Jiang, W.~K.-S. Tang, and G.~Chen}, {\em A state-observer-based
  approach for synchronization in complex dynamical networks}, IEEE
  Transactions on Circuits and Systems I, 53 (2006), pp.~2739--2745.

\bibitem{Kar_93}
{\sc A.~Karr}, {\em Probability}, Springer-Verlag, 2009.

\bibitem{Kim_DeP_15}
{\sc H.~Kim and C.~D. Persis}, {\em Output synchronization of lure-type
  nonlinear systems in the presence of input disturbances}, in 54th IEEE
  Conference on Decision and Control, 2015, pp.~4145 -- 4150.

\bibitem{Liu_Bar_Slo_11}
{\sc Y.~Liu, A.~Barabasi, and J.~Slotine}, {\em Controllability of complex
  networks}, Nature, 473 (2011), pp.~167--173.

\bibitem{Mao_97}
{\sc X.~Mao}, {\em Stochastic Differential Equations and Applications},
  Woodhead Publishing, 1997.

\bibitem{Oks_07}
{\sc B.~Oksendal}, {\em Stochastic Differential Equations: An Introduction with
  Applications (Universitext)}, Springer (New York), 6th~ed., 2007.

\bibitem{Pav_Pog_Wou_Nij}
{\sc A.~Pavlov, A.~Pogromvsky, N.~van~de Wouv, and H.~Nijmeijer}, {\em
  Convergent dynamics, a tribute to {B}oris {P}avlovich {D}emidovich}, Systems
  and Control Letters, 52 (2004), pp.~257--261.

\bibitem{Pec_Car_Joh_Mar_97}
{\sc L.~M. Pecora, T.~L. Carroll, G.~A. Johnson, and D.~J. Mar}, {\em
  Fundamentals in chaotic systems, concepts and applications}, Chaos, 7 (1997),
  pp.~520--543.

\bibitem{DeP_Jay_14}
{\sc C.~D. Persis and B.~Jayawardhana}, {\em On the internal model principle in
  the coordination of nonlinear systems}, IEEE Transactions on Control of
  Network Systems, 1 (2014), pp.~272 -- 282.

\bibitem{Pet_Sri_Tay_Sur_Eck_Bul_15}
{\sc J.~Peters, V.~Srivastava, G.~Taylor, A.~Surana, M.~Eckstein, and
  F.~Bullo}, {\em Human supervisory control of robotic teams: integrating
  cognitive modeling with engineering design}, IEEE Control Systems Magazine,
  (2015), pp.~57--80.

\bibitem{Rat_McK_08}
{\sc R.~Ratcliff and G.~McKoon}, {\em The diffusion decision model: theory and
  data for two-choice decision tasks}, Neural Computation, 20 (2008),
  pp.~873--922.

\bibitem{Roh_76}
{\sc V.~Rohatgi}, {\em An introduction to probability theory and mathematical
  statistics}, John Wiley Sons, 1976.

\bibitem{Rus_diB_09b}
{\sc G.~Russo and M.~di~Bernardo}, {\em How to synchronize biological clocks},
  Journal of Computationa Biology, 16 (2009), pp.~379--393.

\bibitem{Rus_diB_d}
{\sc G.~Russo and M.~di~Bernardo}, {\em Solving the rendezvous problem for
  multi-agent systems using contraction theory}, in Proceedings of the
  International Conference on Decision and Control, 2009, pp.~5821 -- 5826.

\bibitem{Rus_diB_Son_13}
{\sc G.~Russo, M.~di~Bernardo, and E.~D. Sontag}, {\em A contraction approach
  to the hierarchical analysis and design of networked systems}, IEEE
  Transactions on Automatic Control, 58 (2013), pp.~1328--1331.

\bibitem{Sca_Arc_Son_10}
{\sc L.~Scardovi, M.~Arcak, and E.~Sontag}, {\em Synchronization of
  interconnected systems with applications to biochemical networks: an
  input-output approach}, IEEE Transactions on Automatic Control, 55 (2010),
  pp.~1367--1379.

\bibitem{Sha_Arc_15}
{\sc S.~Shafi and M.~Arcak}, {\em Adaptive synchronization of diffusively
  coupled systems}, IEEE Transactions on Control of Network Systems, 2 (2015),
  pp.~131--141.

\bibitem{Sri_Leo_14}
{\sc V.~Srivastava and N.~Leonard}, {\em Collective decision-making in ideal
  networks: the speed-accuracy tradeoff}, IEEE Transactions on Control Of
  Network Systems, 1 (2014), pp.~121--132.

\bibitem{Stu_Sho_12}
{\sc S.~Stidli, E.~Crisostomi, R.~Middleton, and R.~Shorten}, {\em A flexible
  distributed framework for realising electric and plug-in hybrid vehicle
  charging policies}, International Journal of Control, 85 (2012),
  pp.~1130--1145.

\bibitem{Tan_Li_15}
{\sc H.~Tang and T.~Li}, {\em Continuous-time stochastic consensus: stochastic
  approximation and kalman-bucy filtering based protocols}, Automatica, 61
  (2015), pp.~146--155.

\bibitem{Wei_Van_13}
{\sc J.~Wei and A.~van~der Schaft}, {\em Load balancing of dynamical
  distribution networks with flow constraints and unknown in/outflows}, Systems
  \& Control Letters, 62 (2013), pp.~1001 -- 1008.

\bibitem{Wel_Kat_Mot_15}
{\sc D.~Wells, W.~Kath, and A.~Motter}, {\em Control of stochastic and induced
  switchnig in biophysical networks}, Physical Review X, 15 (2015), p.~031036.

\end{thebibliography}
\end{document}